\documentclass[11pt]{article}

\usepackage{amsmath,amssymb}
\usepackage{graphicx}
\usepackage{color}
\usepackage[arrow, matrix, curve]{xy}

\newcommand{\dd}[2]{\frac{\mathrm{d} #1}{\mathrm{d} #2}}
\newcommand{\pdl}[2]{\frac{\partial #1}{\partial #2}}

\newcommand{\mm}{\mathbf{m}}
\newcommand{\nn}{\nabla}
\newcommand{\kk}{\mathbf{k}}

\newcommand{\xx}{\mathbf{x}}
\newcommand{\vv}{\mathbf{v}}

\begin{document}

\par\noindent {\LARGE\bf
Symmetry justification of Lorenz' maximum\\ simplification
\par}
{\vspace{4mm}\par\noindent {\bf Alexander Bihlo~$^\dag$ and Roman O. Popovych~$^\dag \phantom{,}^\ddag$
} \par\vspace{2mm}\par}

{\vspace{2mm}\par\noindent {\it
$^{\dag}$~Faculty of Mathematics, University of Vienna, Nordbergstra{\ss}e 15, A-1090 Vienna, Austria\\
}}
{\noindent \vspace{2mm}{\it
$\phantom{^\dag}$~\textup{E-mail}: alexander.bihlo@univie.ac.at
}\par}

{\vspace{2mm}\par\noindent {\it
$^\ddag$~Institute of Mathematics of NAS of Ukraine, 3 Tereshchenkivska Str., 01601 Kyiv, Ukraine\\
}}
{\noindent \vspace{2mm}{\it
$\phantom{^\dag}$~\textup{E-mail}: rop@imath.kiev.ua
}\par}

{\vspace{7mm}\par\noindent\hspace*{8mm}\parbox{140mm}{\small
In 1960 Edward Lorenz (1917--2008) published a pioneering work on the `maximum simplification' of the barotropic vorticity equation. 
He derived a coupled three-mode system and interpreted it as the minimum core of large-scale fluid mechanics on a `finite but unbounded' domain. 
The model was obtained in a heuristic way, without giving a rigorous justification for the chosen selection of modes. 
In this paper, it is shown that one can legitimate Lorenz' choice by using symmetry transformations of the spectral form of the vorticity equation. 
The Lorenz three-mode model arises as the final step in a hierarchy of models constructed via the component reduction by means of symmetries. 
In this sense, the Lorenz model is indeed the `maximum simplification' of the vorticity equation.
}\par\vspace{7mm}}

\section{Introduction}

Symmetry is one of the most important concepts in numerous branches of modern natural science. 
The exploitation of symmetries of dynamical systems may lead to a more efficient treatment 
of the differential equations describing these systems 
via a reduction of the information that is necessary in order to account for model dynamics.

There are many ways to utilize symmetries of differential equations, including the systematic construction of exact solutions of PDEs, 
determination of conservation laws and construction of mappings that relate or linearize 
differential equations (see e.g.\ \cite{anco02Ay,Olver1993,Ovsiannikov1982}). 
Also, there is some work on symmetries in the study of dynamical systems and bifurcation theory \cite{golu85}, 
giving rise to the study of equivariant dynamical systems. In the present work, we will take a related but somewhat 
different direction to investigate how the Lorenz-1960 model \cite{lor1960} can be derived in a rigorous way.

In this classical work, Lorenz considered the spectral expansion of the barotropic vorticity equation on a torus. In what follows, he sought for 
the minimum system of coupled ordinary first order differential equations for the Fourier coefficients that is necessary to still account for 
the nonlinear interaction of modes. In doing so, he first restricted the range of indices in the infinite Fourier series by the values 
$\{-1,0,1\}$. His crucial step to achieve the maximum simplification was the observation that all but three of these remaining coefficients 
retain their particular values once they are taken. We aim to give a justification of this observation by interpreting it as a condition of 
symmetry. To be more precise, we show that this simplification is possible due to the corresponding spectral counterparts of point symmetry 
transformations of the vorticity equation in physical space. These symmetries are preserved under truncations of the infinite Fourier series and 
are inherited by the spectral set of equations for the Fourier coefficients.

A justification of the selection of modes in finite-mode models by considering inherited symmetries may be potentially useful also in more 
general situations. 
It could provide an additional criterion addressing the important question which modes in the 
reduced model should be retained and which may be neglected. 
In this sense, using induced symmetries may supply special kinds of truncations 
that are designed to preserve e.g.\ the invariants of the parental model also in the truncated dynamics (e.g.\ \cite{thif96}). 
This may lead to more concise and consistent finite-mode representations. 
As stated above, this approach is to be distinguished from the field of equivariant 
dynamics. The main difference is that in the latter usually no exhaustive and rigorous calculations of symmetry groups are given.

The organization of this paper is the following: Section \ref{sec:vort} is devoted to discrete and continuous symmetries of the barotropic vorticity 
equation in a non-rotating reference frame. We expand the vorticity in a double Fourier series and discuss how the symmetries of the equation in 
physical space are induced to symmetry transformations in terms of the Fourier coefficients. In section \ref{sec:discrete} the initial model of 
eight ODEs will be presented amongst a discussion of the induction of the symmetries to this truncated system. 
Finally, in section~\ref{sec:models} subgroups of the whole symmetry group of the truncated spectral vorticity equations 
will be used to derive hierarchies of reduced models describing the evolution of the relevant Fourier coefficients. 
The main result is that the Lorenz-1960 model 
is indeed the maximum simplification of the dynamic equations. 
It is the minimal system that can be obtained by using induced discrete symmetry subgroups of the 
spectral vorticity equation for component reduction.

\section{Symmetries of the barotropic vorticity equation}\label{sec:vort}

The inviscid barotropic vorticity equation in an inertial system in stream function form reads
\begin{equation} \label{vort}
    \pdl{}{t}\nn^2\psi + \pdl{\psi}{x}\pdl{}{y}\nn^2\psi - \pdl{\psi}{y}\pdl{}{x}\nn^2\psi = 0,
\end{equation}
where $\psi$ is the stream function generating two-dimensional nondivergent flow in the $(x,y)$-plane. It states the individual or Lagrangian conservation of the vorticity
\begin{equation}
    \zeta = \nn^2\psi \nonumber.
\end{equation}
Eqn. (\ref{vort}) possesses the eight-element group of discrete symmetries, generated by the elements
\begin{gather*} 
e_1\colon\ (x,y,t,\psi) \to (x,-y,t,-\psi)\\
e_2\colon\ (x,y,t,\psi) \to (-x,y,t,-\psi)\\
e_3\colon\ (x,y,t,\psi) \to (x,y,-t,-\psi).
\end{gather*}
Note that these transformations are involutive (i.e.\ $e_i^2 = 1$, $i=1,2,3$) 
and, moreover, they commute (i.e.\ $e_ie_j = e_je_i$, $i,j = 1,2,3$, $i\ne j$). 
Each $e_i$ generates a copy of $\mathbb{Z}_2$, the cyclic group of order 2. 
The group of all discrete symmetry transformations of the vorticity equation may then be written as $\mathbb{Z}_2\oplus\mathbb{Z}_2\oplus\mathbb{Z}_2$.

For sake of completeness, we also present the basis operators of the maximal Lie invariance algebra of (\ref{vort}), 
which were computed using the program LIE by A.~Head~\cite{Head}.
They read
\begin{align}
    &\vv_t = \pdl{}{t}& &  \vv_u = tx\pdl{}{y}-ty\pdl{}{x}+\frac{1}{2}\left(x^2+y^2\right)\pdl{}{\psi}\nonumber \\
    &\vv_r = x\pdl{}{y}-y\pdl{}{x}& & \mathcal{Z}(h) = h(t)\pdl{}{\psi}\nonumber \\
    &\mathcal{X}_1(f) = f(t)\pdl{}{x} - yf'(t)\pdl{}{\psi}& & \mathcal{X}_2(g) = g(t)\pdl{}{y} + xg'(t)\pdl{}{\psi}\nonumber \\
    &\mathcal{D}_1 = x\pdl{}{x}+y\pdl{}{y}+2\psi\pdl{}{\psi}& & \mathcal{D}_2 = t\pdl{}{t}-\psi\pdl{}{\psi},\nonumber
\end{align}
where $f$, $g$ and $h$ are arbitrary smooth functions of~$t$.
Thus, in addition to the discrete symmetries $e_i$, $i=1,2,3$, the vorticity equation (\ref{vort}) possesses 
the time translations (generated by $\vv_t$), 
the rotations with constant velocities and on constant angles ($\vv_u$ and $\vv_r$, respectively), 
the gauging of stream function with arbitrary summands depending in~$t$ ($\mathcal{Z}(h)$), 
translatory motions with arbitrary (nonconstant) velocities ($\mathcal{X}_1(f)$ and $\mathcal{X}_2(g)$) 
and (separate) scaling of the space and time variables ($\mathcal{D}_1$ and $\mathcal{D}_2$). 
In contrast to the other basis operators, the operator $\vv_u$ has no counterpart in the three-dimensional case 
and leads to nonlocal transformations in terms of the fluid velocity and the pressure. 
This singularity, from the symmetry point of view, of the two-dimensional vorticity equations in terms of the stream function 
was first observed by Berker~\cite{Berker1963} and later re-opened (c.f.~\cite{Andreev&Rodionov1988}). 
In this paper we simultaneously use discrete symmetries and some of transformations from the continuous symmetry group.

We now expand the vorticity in a double Fourier series on the torus,
\[
    \zeta = \sum_{\mm}c_{\mm}\exp(i\hat\mm\cdot\xx), \qquad 
    \xx=x\mathbf i+y\mathbf j,\quad \mm=m_1\mathbf i+m_2\mathbf j,\quad \hat\mm=m_1k\mathbf i+m_2l\mathbf j,
\]
where $k$ and $l$ are nonzero constants, $\mathbf i=(1,0,0)^{\rm T}$, $\mathbf j=(0,1,0)^{\rm T}$, 
$m_1$ and $m_2$ run through the set of integers and the coefficient $c_{00}$ vanishes. 
Inserting this expansion in the vorticity equation (\ref{vort}) gives its spectral form~\cite{lor1960}:
\begin{equation} \label{specvort}
    \dd{c_\mm}{t} = -\sum_{\mm'\ne\mathbf{0}}\frac{c_{\mm'}c_{\mm-\mm'}}{\hat\mm'{}^2}\left(\kk\cdot[\hat\mm'\times\hat\mm]\right),
\end{equation}
where $\kk=(0,0,1)^{\rm T}$. 
The transformations $e_i$ induce transformations of the Fourier coefficients. 
Thus, in spectral terms the action $\tilde\zeta(x, y, t) = -\zeta(x,-y,t)$ of $e_1$ on $\zeta$ has the form 
\begin{gather*}
    \sum_\mm \tilde c_\mm\exp(i\hat\mm\cdot\xx) = -\sum_\mm c_\mm\exp(i(m_1kx-m_2ly))= -\sum_\mm c_{m_1,-m_2}\exp(i(m_1kx+m_2ly)),
\end{gather*}
upon changing the summation over $m_2$. 
Hence we have $\tilde c_{m_1m_2} = -c_{m_1,-m_2}$ as a consequence of the $x$-reflection $e_1$. 
Finally, these and similar computations give the transformations
\begin{align*}
    e_1\colon& \quad c_{m_1m_2} \to -c_{m_1,-m_2} \\
    e_2\colon& \quad c_{m_1m_2} \to -c_{-m_1m_2} \\
    e_3\colon& \quad c_{m_1m_2} \to -c_{m_1m_2}, \quad t \to -t.
\end{align*}
It is obvious that the induced transformations for the coefficients $c_{\mm}$ are symmetries of system~(\ref{specvort}).

Some of continuous symmetries of the vorticity equation also induce well-defined symmetries of system (\ref{specvort}) 
which are applicable within the framework of our approach. 
Thus, the rotation on the angle~$\pi$ coincides with $e_1e_2$. 
(In fact, only one of the reflections $e_1$ and $e_2$ is independent up to continuous symmetries.) 
More nontrivial examples are given by the space translations, which induce the transformations 
\begin{align*}
    p_\varepsilon\colon& \quad c_{m_1m_2} \to e^{im_1k\varepsilon}c_{m_1m_2} \\
    q_\varepsilon\colon& \quad c_{m_1m_2} \to e^{im_2l\varepsilon}c_{m_1m_2}.
\end{align*}
For the values $\varepsilon=\pi/k$ and $\varepsilon=\pi/l$, respectively we have 
$p_{\pi/k}\colon c_{m_1m_2} \to (-1)^{m_1}c_{m_1m_2}$ and 
$q_{\pi/l}\colon c_{m_1m_2} \to (-1)^{m_2}c_{m_1m_2}$. 
For the sake of brevity, we will set $\hat p = p_{\pi/k}$ and $\hat q = q_{\pi/l}$.
It is easy to see that the transformations $\hat p$ and $\hat q$ are involutive and commutes with $e_1$, $e_2$ and $e_3$.

\section{Discrete symmetries of the truncated system}\label{sec:discrete}

The restriction of the range of indices in (\ref{specvort}) by $\{-1,0,1\}\times\{-1,0,1\}$ leads to the following eight-mode model related to the vorticity equation:
\begin{align}\label{system}
    \dd{c_{11}}{t} &= \left(\frac{1}{l^2}-\frac{1}{k^2}\right)klc_{10}c_{01}& \dd{c_{10}}{t} &= \left(\frac{1}{k^2+l^2}-\frac{1}{l^2}\right)kl\left[c_{11}c_{0,-1} - c_{1,-1}c_{01} \right] \nonumber\\
    \dd{c_{1,-1}}{t} &= \left(\frac{1}{k^2}-\frac{1}{l^2}\right)klc_{10}c_{0,-1}&   \dd{c_{01}}{t} &= \left(\frac{1}{k^2+l^2}-\frac{1}{k^2}\right)kl\left[c_{10}c_{-11}-c_{11}c_{-10}\right] \nonumber\\
    \dd{c_{-1,-1}}{t} &= \left(\frac{1}{l^2}-\frac{1}{k^2}\right)klc_{-10}c_{0,-1}&   \dd{c_{-10}}{t} &= \left(\frac{1}{k^2+l^2}-\frac{1}{l^2}\right)kl\left[c_{-1,-1}c_{01} - c_{0,-1}c_{-11} \right] \nonumber\\
    \dd{c_{-11}}{t} &= \left(\frac{1}{k^2}-\frac{1}{l^2}\right)klc_{-10}c_{01}&   \dd{c_{0,-1}}{t} &= \left(\frac{1}{k^2+l^2}-\frac{1}{k^2}\right)kl\left[c_{-10}c_{1,-1} - c_{10}c_{-1,-1}\right]. 
\end{align}
System (\ref{system}) consists of first-order ordinary differential equations. 
Hence the problem on description of its point symmetries is even more difficult than its complete integration (c.f.\ \cite{Olver1993}). 
At the same time, some symmetries of (\ref{system}) are in fact known since they are induced by symmetries of the vorticity equation. 
In particular, the symmetric truncation guarantees that the resulting system (\ref{system}) 
inherits the discrete symmetries of the initial system (\ref{specvort}), induced by $e_1$, $e_2$, $e_3$ and their compositions. 
(This is an argument justifying such kind of truncation.) 
Any truncation also preserves the symmetries $p_\varepsilon$ and $q_\varepsilon$, in particular, $\hat p$ and $\hat q$.

Induced transformations are of crucial importance for deriving the Lorenz system.

\section{Component reduction of the truncated system}\label{sec:models}

The transformations $e_1$, $e_2$, $\hat p$ and $\hat q$ and their compositions act only on dependent variables 
and, therefore, can be used for component reductions of system (\ref{system}) within our approach.%
\footnote{See the appendix for a depiction of these transformations.} 
The technique applied is similar to that developed for invariant solutions without transversality 
(c.f.\ \cite{Anderson&Fels&Torre2000,Ovsiannikov1982}). 
However, it is not apparent that the element $e_3$ can be used for this purpose. 
We thus restrict ourself to the group $G\simeq\mathbb{Z}_2\oplus\mathbb{Z}_2\oplus\mathbb{Z}_2\oplus\mathbb{Z}_2$
generated by $e_1$, $e_2$, $\hat p$ and $\hat q$.
We will apply the following subgroups of~$G$:
\begin{gather*}
    S_1 = \{1,e_1\},\quad S'_1 = \{1,e_2\},\quad 
    S_2 = \{1,\hat p\},\quad S'_2 = \{1,\hat q\},\\ 
    S_3 = \{1,\hat p e_2\},\quad S'_3 = \{1,\hat q e_1\},\quad 
    S_4 = \{1,\hat p \hat q\},\\ 
    S_5 = \{1,\hat p e_1\},\quad S'_5 = \{1,\hat q e_2\},\quad
    S_6 = \{1,e_1e_2\},\\
    S_7 = \{1,\hat p e_1e_2\},\quad S'_7 = \{1,\hat q e_1e_2\},\quad
    S_8 = \{1,\hat p \hat q  e_1\},\quad S'_8 = \{1,\hat p \hat q e_2\},\\
    S_9 = \{1,\hat p \hat q e_1e_2\},\quad S_{10} = \{1,\hat p \hat q e_1,\hat p \hat q e_2,e_1e_2\},\\    
    S_{11} = \{1,\hat p e_1,\hat q e_1e_2,\hat p \hat q e_2\},\quad S'_{11} = \{1,\hat q e_2,\hat p e_1e_2,\hat p \hat q e_1\},\\    
    S_{12} = \{1,\hat p e_1,\hat q e_2,\hat p \hat q e_1e_2\}.    
\end{gather*}
By 1 we denote the identical transformation.
There are yet other subgroups of $G$ but as we will see they do not lead to nontrivial reduced systems.

Note that the subgroup $S'_1$ will result in the same reduced system as $S_1$ (up to the re-notation $(x,k) \leftrightarrow (y,l)$). 
A similar remark holds also for all the subgroup marked by prime, 
so we only have to consider reductions with respect to the subgroups without prime. 
This should be done subsequently.

\subsection{Trivial reductions}

The tuple $(c_{ij})$ is invariant with respect to the transformation $e_1$ if and only if the following identifications hold:
\[
    c_{-1,-1}=-c_{-11},\quad  c_{0,-1}=-c_{01},\quad c_{1,-1}=-c_{11},\quad c_{10}=-c_{10},\quad c_{-10}=-c_{-10}. 
\]
The last two conditions require that $c_{10}= c_{-10} = 0$. 
Inserting these identifications in (\ref{system}) leads to a trivial system only. 
Hence, the subgroup $S_1$ cannot be used for a component reduction of~(\ref{system}). 
As a consequence, no subgroup of $G$ that contains the transformation $e_1$ can be used for this purpose. 

The same statement is true for the transformations $e_2$, $\hat p$, $\hat q$, $\hat p e_2$, $\hat q e_1$ and $\hat p \hat q$.
Finally, any subgroup of $G$ containing one of the elements 
$e_1$, $e_2$, $\hat p$, $\hat q$, $\hat p e_2$, $\hat q e_1$ or $\hat p \hat q$
gives a trivial reduction.

\subsection{Reductions in three components}

The tuple $(c_{ij})$ is invariant under the subgroup~$S_8$ generated by the transformation $\hat p \hat q  e_1$ if and only if the following equalities hold:
\[
    c_{0,-1} = c_{01},\quad c_{-1,-1} = -c_{-11},\quad c_{1,-1} = -c_{11}.
\]
This transformation allows us to reduce (\ref{system}) to the nontrivial five-component system
\begin{align}\label{hiddensystem}
      \dd{c_{10}}{t} &= 2\left(\frac{1}{k^2+l^2}-\frac{1}{l^2}\right)klc_{11}c_{01} & \dd{c_{11}}{t} &= \left(\frac{1}{l^2}-\frac{1}{k^2}\right)klc_{10}c_{01} \nonumber \\
        \dd{c_{-10}}{t} &= -2\left(\frac{1}{k^2+l^2}-\frac{1}{l^2}\right)klc_{-11}c_{01} & \dd{c_{-11}}{t} &= \left(\frac{1}{k^2}-\frac{1}{l^2}\right)klc_{-10}c_{01} \\
    \dd{c_{01}}{t} &= \left(\frac{1}{k^2+l^2}-\frac{1}{k^2}\right)kl\left[c_{10}c_{-11}-c_{11}c_{-10}\right].\nonumber
\end{align}

The subgroup~$S_5$ leads to a similar reduction under equating 
\[
    c_{0,-1} = -c_{01},\quad c_{-1,-1} = c_{-11},\quad c_{1,-1} = c_{11}.
\]

\subsection{Reductions in four components}

Utilizing the subgroup $S_6=\{1,e_1e_2\}$ by means of similar consideration as in the previous sections leads 
to the following nontrivial reduced system of (\ref{system}) after equating 
$c_{-11}=c_{1,-1}$, $c_{-1,-1}=c_{11}$, $c_{0,-1}=c_{01}$ and $c_{-10}=c_{10}$:
\begin{align} 
    \dd{c_{11}}{t} &= \left(\frac{1}{l^2}-\frac{1}{k^2}\right)klc_{10}c_{01} & 
    \dd{c_{10}}{t} &= \left(\frac{1}{k^2+l^2}-\frac{1}{l^2}\right)kl[c_{11}-c_{1,-1}]c_{01} \nonumber \\
    \dd{c_{1,-1}}{t} &= \left(\frac{1}{k^2}-\frac{1}{l^2}\right)klc_{10}c_{01} & \dd{c_{01}}{t} 
    &= \left(\frac{1}{k^2+l^2}-\frac{1}{k^2}\right)kl\left[c_{1,-1} - c_{11}\right]c_{10}.\label{system2}
\end{align}

Similar four-component reductions are also given by the non-primed subgroups 
$S_7$ ($c_{-11}=-c_{1,-1}$, $c_{-1,-1}=-c_{11}$, $c_{0,-1}=c_{01}$ and $c_{-10}=-c_{10}$)
and $S_9$ ($c_{-11}=c_{1,-1}$, $c_{-1,-1}=c_{11}$, $c_{0,-1}=-c_{01}$ and $c_{-10}=-c_{10}$).

\subsection{Maximal reduction: subgroup $\boldsymbol{S_{10}}$}

Finally, let us derive the reduced system associated with the subgroup~$S_{10}$. 
This reduced system coincides with the Lorenz-1960 model. 
Since all transformations mutually commute, there are three equivalent ways for the reduction.
The first way is to perform the reduction in a single step from system (\ref{system}) using the subgroup~$S_{10}$.
Because the system (\ref{system2}) possesses the symmetry transformation induced by $\hat p \hat q  e_1$, 
we can also start with (\ref{system2}) and compute the induced component reduction. 
Alternatively, it is possible start with (\ref{hiddensystem}) and apply the transformation induced by $e_1e_2$. 
This will give the same reduced system associated with the subgroup~$S_{10}$. 

The subgroups $S_{11}$ and $S_{12}$ lead to similar reductions connected with the Lorenz' reduction via 
transformations generated by $p_{\pi/{2k}}$ and $q_{\pi/{2l}}$. 
Therefore, up to symmetries of (\ref{system}) induced by the symmetries of the vorticity equation, 
there is a unique nontrivial reduction in five components using subgroups of~$G$.

However, let us note before that we have already employed the transformation \[(x,y,t,\psi)\to (-x,-y,t,\psi)\] which accounts for the identification $c_{-\mm} = c_{\mm}$. Observe that the connection between the complex and real Fourier coefficients is given by
\[
    a_\mm = c_\mm + c_{-\mm}, \qquad b_\mm = i(c_\mm-c_{-\mm}).
\] 
Hence we can already set

\begin{equation}
    c_\mm = \frac{1}{2}a_\mm. \nonumber
\end{equation}
This explains the first observation by Lorenz \cite{lor1960}: 
If the imaginary parts of the Fourier coefficients vanish initially, they will vanish for all times. 
Mathematically, this is justified since we have used the symmetry transformation group $S_6$.

Now employing the transformation $\hat p \hat q  e_1$ at this stage of simplification, we have the identification
\begin{equation}
    c_{1,-1} = -c_{11} \quad \Leftrightarrow \quad a_{1,-1} = -a_{11}. \nonumber
\end{equation}
This is the second observation by Lorenz \cite{lor1960}, namely that if $a_{1,-1} = -a_{11}$ initially, then the equality will hold for all times. 
That is, Lorenz heuristically discovered the symmetry transformation $\hat p \hat q  e_1$ to obtain his famous finite-mode model.

It is straightforward to see that the resulting system leads to the Lorenz system upon setting $A = a_{01}, F = a_{10}, G = a_{1,-1}$:
\begin{align}
    \dd{A}{t} &= -\left(\frac{1}{k^2}-\frac{1}{k^2+l^2}\right)klFG \nonumber \\[1ex]
    \dd{F}{t} &= \left(\frac{1}{l^2}-\frac{1}{k^2+l^2}\right)klAG \nonumber \\[1ex]
    \dd{G}{t} &= -\frac{1}{2}\left(\frac{1}{l^2}-\frac{1}{k^2}\right)klAF. \nonumber
\end{align}
The Lorenz-1960 system is thus the result of invariance of the system (\ref{system}) under the subgroup $S_{10}$. 
Hence it is truly the maximum simplification that can be obtained within our symmetry framework. 
We have exhausted all the symmetry groups consisting of chosen discrete symmetry transformations and leading to \emph{nontrivial} reduced systems.

\section*{Acknowledgments}

The authors are grateful to the referees for their remarks. The authors thank Michael Hantel and Leopold Haimberger for helpful discussions. The research of ROP was supported by the Austrian Science Fund (FWF), START-project Y237 and stand-alone project P20632. AB is a recipient of a DOC-fellowship of the Austrian Academy of Science.

\section*{Appendix: Depiction of symmetry transformations}\label{sec:app}

For the readers' convenience, we graphically illustrate some of the symmetry transformations used for the component reduction. 
An arrow between two components means the identification of the respective coefficients. 
If there is a minus sign next to an arrow, the corresponding identification is up to minus. 
For any arrow starting and ending in the same coefficient, the identification up to minus means that this coefficient vanishes. 
The coefficients remained in the finite-mode models after the identification
are displayed in bold.

\noindent
\begin{figure}[htdp]
\centering
\begin{minipage}[t]{0.4\textwidth}
\centering
\begin{xy}
\xymatrix{
      \boldsymbol{c_{-11}}\ar@{<->}@/_0.5cm/[dd]_{-} & \boldsymbol{c_{01}}\ar@{<->}[dd]^{-} & \boldsymbol{c_{11}}\ar@{<->}@/^0.5cm/[dd]^{-}  \\
      c_{-10}\ar@(dr,ur)[]_{-} &  & c_{10}\ar@(dl,ul)[]^{-}  \\
      c_{-1,-1}  &   c_{0,-1}  & c_{1,-1}
  }
\end{xy}
\vspace{0.5cm}
\centerline{Symmetry transformation $e_1$\quad}
\end{minipage}
\begin{minipage}[t]{0.4\textwidth}
\centering
\begin{xy}
\  \xymatrix{
      \boldsymbol{c_{-11}}\ar@{<->}@/_0.5cm/[dd]_{-} & \boldsymbol{c_{01}}\ar@{<->}[dd] & \boldsymbol{c_{11}}\ar@{<->}@/^0.5cm/[dd]^{-}  \\
      \boldsymbol{c_{-10}}\ar@(dr,ur)[]  &  & \boldsymbol{c_{10}}\ar@(dl,ul)[]  \\
      c_{-1,-1}    &   c_{0,-1} & c_{1,-1}
  }
\end{xy}
\vspace{0.5cm}
\centerline{Symmetry transformation $\hat p \hat q  e_1$}
\end{minipage}

\vspace{10mm}

\centering
\begin{minipage}[t]{0.4\textwidth}
\begin{center}
\begin{xy}
\xymatrix{
      c_{-11}\ar@{<->}@/^0.5cm/[rr]^{-} & c_{01}\ar@(dl,dr)[]^{-} & \boldsymbol{c_{11}}  \\
      c_{-10}\ar@{<->}[rr]_{-} &  & \boldsymbol{c_{10}}  \\
      c_{-1,-1}\ar@{<->}@/_0.5cm/[rr]_{-} & c_{0,-1}\ar@(ul,ur)[]_{-}  & \boldsymbol{c_{1,-1}}
  }
\end{xy}
\vspace{0.3cm}
Symmetry transformation $e_2$\qquad\quad\null 
\end{center}
\end{minipage}
\begin{minipage}[t]{0.4\textwidth}
\begin{center}
\begin{xy}
\ \quad\xymatrix{
      c_{-11}\ar@{<->}@/^0.5cm/[rr]^{-} & \boldsymbol{c_{01}}\ar@(dl,dr)[] & \boldsymbol{c_{11}}  \\
      c_{-10}\ar@{<->}[rr] &  & \boldsymbol{c_{10}}  \\
      c_{-1,-1}\ar@{<->}@/_0.5cm/[rr]_{-} & \boldsymbol{c_{0,-1}}\ar@(ul,ur)[]  & \boldsymbol{c_{1,-1}}
  }
\end{xy}
\vspace{0.3cm}
Symmetry transformation $\hat p \hat q  e_2$
\end{center}
\end{minipage}

\vspace{10mm}

\begin{minipage}[t]{0.4\textwidth}
\begin{center}
\begin{xy}
\xymatrix{
      \boldsymbol{c_{-11}}\ar@(dl,dr)[] & c_{01}\ar@(dl,dr)[]^{-} & \boldsymbol{c_{11}}\ar@(dl,dr)[]\\
      c_{-10}\ar@(dr,ur)[]_{-} & & c_{10}\ar@(dl,ul)[]^{-}  \\
      \boldsymbol{c_{-1,-1}}\ar@(ul,ur)[] & c_{0,-1}\ar@(ul,ur)[]_{-} & \boldsymbol{c_{1,-1}}\ar@(ul,ur)[]
  }
\end{xy}
\vspace{0.5cm}
Symmetry transformation $\hat p \hat q$\qquad\quad\null
\end{center}
\end{minipage}
\begin{minipage}[t]{0.4\textwidth}
\begin{center}
\begin{xy}
\ \quad\xymatrix{
      c_{-11}\ar@{<->}[ddrr] & \boldsymbol{c_{01}}\ar@<-0.3mm>@{<->}[dd] & \boldsymbol{c_{11}}\ar@{<->}[ddll]  \\
      c_{-10}\ar@{<->}[rr] &  & \boldsymbol{c_{10}}\\
      c_{-1,-1} & c_{0,-1} & \boldsymbol{c_{1,-1}}
  }
\end{xy}
\vspace{0.5cm}
Symmetry transformation $e_1e_2$
\end{center}
\end{minipage}

\end{figure}

\end{document}